\documentclass[aps,axodraw,twocolumn,prl]{revtex4}
\usepackage{fancyhdr}
\usepackage{fancybox}
\usepackage{graphicx}
\usepackage{color}
\usepackage{soul}
\usepackage{latexsym}

\def\sig{\sigma}
\def\nobs{n_{\rm obs}}
\def\nexp{n_{\rm exp}}

\def\cbb{C_{eff}^{2b}}
\def\cbbbb{C_{eff}^{4b}}

\def\ma{m_a}
\def\mh{m_h}
\def\hi{h_1}

\def\ai{a_1}
\def\mhi{m_{h_1}}

\def\mai{m_{a_1}}

\def\mtau{m_\tau}

\def\lam{\lambda}
\def\kap{\kappa}
\def\alam{A_\lambda}
\def\akap{A_\kappa}

\def\mh{m_h}

\def\h{h}
\def\mh{m_{h}}

\def\hbar{\overline h}
\def\lam{\lambda}

\def\mx{M_X}

\def\mz{m_Z}

\def\hi{h_i^0}
\def\mhi{m_{\hi}}

\def\h{h}
\def\mh{m_{\h}}

\def\lam{\lambda}

\def\wtil{\widetilde}

\def\tauptaum{\tau^+\tau^-}

\def\lsim{\mathrel{\raise.3ex\hbox{$<$\kern-.75em\lower1ex\hbox{$\sim$}}}}
\def\gsim{\mathrel{\raise.3ex\hbox{$>$\kern-.75em\lower1ex\hbox{$\sim$}}}}
\def\ifmath#1{\relax\ifmmode #1\else $#1$\fi}

\def\vev#1{\langle #1 \rangle}
\def\lam{\lambda}

\def\mhi{m_{h_1}}

\def\susy{{\rm SUSY}}

\def\hsm{h_{\rm SM}}

\def\tanb{\tan\beta}

\def\mb{m_b}
\def\mz{m_Z}

\def\mx{M_X}

\def\cnone{\wt\chi^0_1}

\def\cntwo{\wt\chi^0_2}

\def\mcnone{m_{\cnone}}
\def\mcntwo{m_{\cntwo}}

\def\wt{\widetilde}

\def\MPL #1 #2 #3 {{\sl Mod.~Phys.~Lett.}~{\bf#1} (#3) #2}
\def\NPB #1 #2 #3 {{\sl Nucl.~Phys.}~{\bf #1} (#3) #2}
\def\PLB #1 #2 #3 {{\sl Phys.~Lett.}~{\bf #1} (#3) #2}
\def\PR #1 #2 #3 {{\sl Phys.~Rep.}~{\bf#1} (#3) #2}
\def\PRD #1 #2 #3 {{\sl Phys.~Rev.}~{\bf #1} (#3) #2}
\def\PRL #1 #2 #3 {{\sl Phys.~Rev.~Lett.}~{\bf#1} (#3) #2}
\def\RMP #1 #2 #3 {{\sl Rev.~Mod.~Phys.}~{\bf#1} (#3) #2}
\def\ZPC #1 #2 #3 {{\sl Z.~Phys.}~{\bf #1} (#3) #2}
\def\IJMP #1 #2 #3 {{\sl Int.~J.~Mod.~Phys.}~{\bf#1} (#3) #2}
\def\NIM #1 #2 #3 {{\sl Nucl.~Inst.~and~Meth.}~{\bf#1} {#3} #2}

\def\lam{\lambda}
\def\br{B}
\def\tauptaum{\tau^+\tau^-}

\def\gam{\gamma}

\def\anti{\overline}
\def\epem{e^+e^-}

\def\anti{\overline}

\def\ai{a_1}

\def\mai{m_{\ai}}

\def\fbi{~{\rm fb}^{-1}}

\def\gev{~{\rm GeV}}
\def\tev{~{\rm TeV}}

\def\mb{m_b}

\def\hi{\h_1}

\def\mhi{m_{\hi}}

\newcommand{\nc}{\newcommand}
\nc{\beq}{\begin{equation}}   \nc{\eeq}{\end{equation}}
\nc{\bea}{\begin{eqnarray}}   \nc{\eea}{\end{eqnarray}}
\nc{\baa}{\begin{array}}      \nc{\eaa}{\end{array}}
\nc{\bit}{\begin{itemize}}    \nc{\eit}{\end{itemize}}
\nc{\ben}{\begin{enumerate}}  \nc{\een}{\end{enumerate}}
\nc{\bce}{\begin{center}}     \nc{\ece}{\end{center}}
\def\beqa{\begin{eqnarray}}
\def\eeqa{\end{eqnarray}}
\def\bed{\begin{description}}
\def\eed{\end{description}}

\def\mhi{m_{h_1}}

\def\tanb{\tan\beta}

\def\fbi{~{\rm fb^{-1}}} 
\def\simle{
    \mathrel{\rlap{\raise 0.511ex 
        \hbox{$<$}}{\lower 0.511ex \hbox{$\sim$}}}}

\def\slashchar#1{\setbox0=\hbox{$#1$}           
   \dimen0=\wd0                                 
   \setbox1=\hbox{/} \dimen1=\wd1               
   \ifdim\dimen0>\dimen1                        
      \rlap{\hbox to \dimen0{\hfil/\hfil}}      
      #1                                        
   \else                                        
      \rlap{\hbox to \dimen1{\hfil$#1$\hfil}}   
      /                                         
   \fi}

\def\lam{\lambda}

\def\ls#1{\ifmath{_{\lower1.5pt\hbox{$\scriptstyle #1$}}}}
\def\lss#1{\ifmath{^{\,\lower2.5pt\hbox{$\scriptstyle #1$}}}}

\tighten \preprint{UCD-HEP-???}
\begin{document}
\title{\Large\bf Consistency of LEP $Z+b$-jets excess with an 
{\boldmath $h\to aa$} Decay Scenario and Low-Fine-Tuning NMSSM Models
}
\author{
Radovan Derm\' \i \v sek$^*$ and John F. Gunion$^\dagger$}
\address{
$^*$School of Natural Sciences, Institute for Advanced Study, Princeton,
NJ 08540 \break
$^\dagger$Department of Physics, University of California at Davis, Davis, CA 95616} 
\begin{abstract}
We examine the  LEP limits for the $Zh\to Z+b$'s
final state and find that the excess of observed
events for $\mh\sim 100\gev$ correlates well with
there being an $\mh\sim 100\gev$ Higgs boson with SM-like $ZZh$ coupling
that decays partly via $h\to b\anti b+\tauptaum$ 
[with $\br(h\to b\anti b)\sim 0.08$] but dominantly via $h\to aa$
[with $\br(h\to aa)\sim 0.9$], where $\ma<2\mb$ so that $a\to
\tauptaum$ (or light quarks and gluons) decays are dominant.
Scenarios of precisely this type arise
in the Next-to-Minimal Supersymmetric Model for parameter choices
yielding the lowest possible fine-tuning.
\end{abstract}
\maketitle
\thispagestyle{empty}

\begin{figure}[h!]
\centerline{\includegraphics[width=2.4in,angle=90]{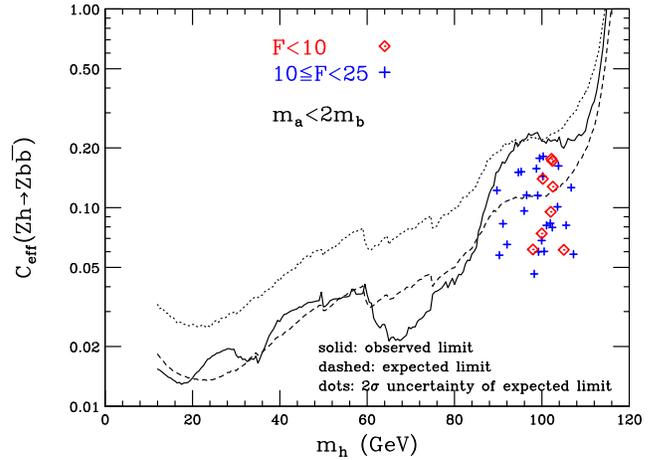}}
\vspace*{-.1in}
\caption{Expected and observed 95\% CL limits on
  $\cbb=[g_{ZZh}^2/g_{ZZ\hsm}^2]\br(h\to b\anti b)$ from
  Ref.~\cite{oldleplimits} are shown vs. $\mh$.  Also plotted are the
  predictions for NMSSM parameter choices in our fixed $\tanb=10$,
  $M_{1,2,3}(\mz)=100,200,300\gev$ scan that give
  fine-tuning measure $F<25$ and $\mai<2\mb$ and that are consistent
  with the preliminary LHWG analysis code.}
\label{zbblimits}
\vspace*{-.3in}
\end{figure}

LEP has placed strong constraints on $Zh$ production where $h$ decays
primarily to $b$-quarks. The limits on
$\cbb=[g_{ZZh}^2/g_{ZZ\hsm}^2]\br(h\to b\anti b)$ are shown in
Fig.~\ref{zbblimits}~\cite{oldleplimits}.  These limits rule out a 
Standard Model (SM)
Higgs boson with $\mh\lsim 114\gev$ at 95\% CL.  However, the plot
also exhibits the well-known excess of observed vs. expected $\cbb$
limits for a test Higgs mass of $\mh\sim 100\gev$.  This excess is
particularly apparent in the $1-CL_b$ result (Fig.~7 of
\cite{oldleplimits}) obtained after combining all four LEP
experiments.  Various interpretations of this excess in terms of a
non-SM Higgs sector have been suggested~\cite{newleplimits,Drees:2005jg}.  In
this letter, we point out that this excess is consistent with a
scenario in which the Higgs boson has SM-like $ZZh$ coupling, but has
reduced $\br(h\to b\anti b)$ by virtue of the presence of $h$ decays
to a pair of lighter Higgs bosons, $h\to aa$, where we use the
notation $a$ appropriate to the NMSSM cases of interest where $a$ is a
CP-odd Higgs boson. (For a generic two-Higgs doublet model with or
without extra singlets, the light Higgs could alternatively be
CP-even, or if CP-violation is present, of mixed CP-nature.) As we
shall discuss, the $a$ must not have significant $\br(a\to b\anti b)$.
The most naturally consistent possibility is $\ma<2\mb$ so that $a\to
\tauptaum$ or light quarks and gluons.  We then emphasize that 
parameter choices for the Next to Minimal Supersymmetric Model (NMSSM) 
that yield the smallest possible fine-tuning 
typically predict precisely this kind of scenario. 

The limits on $\cbb$ apply when the only decays of the $h$ to
$b$-quarks are direct, $h\to b\anti b$. If the only decays of $h$ to
$b$'s were via $Zh\to Zaa\to Zb\anti b b\anti b$, alternative
constraints~\cite{newleplimits} on
$\cbbbb=[g_{ZZh}^2/g_{ZZ\hsm}^2]\br(h\to Zaa)[\br[a \to b\anti b)]^2$
would apply. Again, $\cbbbb$ is above the background Monte Carlo
predictions, in particular for $\mh\gsim 105\gev$ and $\ma\sim 30\div
45\gev$. However, since in the LEP analyzes for the $Z2b$ and $Z4b$
final states the selected candidates are correlated, the $Z2b$ and
$Z4b$ excesses cannot be treated as being independent if the Higgs
boson decays to both kinds of final state. In particular, a model for
which the predictions for $\cbb$ and $\cbbbb$ are individually allowed
but both are close to their respective observed exclusion limits is
actually ruled out because the combined $Z2b$ and $Z4b$ rates give too
large a net $Z+b$'s event rate~\cite{bechtle}.  The
escape we propose is that $\ma<2\mb$ so that $a\to \tauptaum$ (or
$a\to$ light quarks and gluons if $\ma<2\mtau$) is dominant, implying
that $Zh\to Zaa$ does not lead to a $Z+b$'s final state. If the $ZZh$
coupling is full SM strength, then $\mh\sim 100\gev$ with $\br(h\to
b\anti b)\sim 0.08$ and $\br(h\to aa)\sim 0.9$ fits the observed $Z2b$
excess nicely.  Meanwhile, there are no current limits on the $Zh\to
Zaa\to Z\tauptaum\tauptaum$ final state for $\mh\gsim 87\gev$
\cite{newleplimits}.  Further, since the $ZZh$ and $WWh$ couplings are
very SM-like such a model gives excellent agreement with precision
electroweak data.

We are particularly led to the above interpretation of LEP data as a
result of our successful search \cite{Dermisek:2005ar} for NMSSM
parameter choices for which fine-tuning of the model with respect to
unification scale parameters is absent.  We found that the NMSSM
parameters that yield small fine-tuning typically
predict an $\hi$ and $\ai$ with characteristics similar to and
sometimes exactly matching the requirements
described above. In contrast, in the CP-conserving Minimal
Supersymmetric Model (MSSM), $h\to b\anti b$
decays are dominant and all parameter choices 
consistent with LEP limits on $\mh$ are such that the
fine-tuning and hierarchy problems are severe.

The NMSSM is an extremely attractive model \cite{allg}.
First, it provides a very elegant solution to the $\mu$ problem of the
MSSM via the introduction of a singlet superfield $\widehat{S}$.  For
the simplest possible scale invariant form of the superpotential, the
scalar component of $\widehat{S}$ naturally acquires a vacuum
expectation value of the order of the \susy\ breaking scale, giving
rise to a value of $\mu$ of order the electroweak scale. The NMSSM is
the simplest supersymmetric extension of the standard model in which
the electroweak scale originates from the \susy\ breaking scale only.
Hence, the NMSSM deserves very serious consideration.

 Apart from the usual quark and lepton Yukawa
couplings, the scale invariant superpotential of the NMSSM is 
$
W=\lambda \widehat{S} \widehat{H}_u \widehat{H}_d + \frac{\kappa}{3}
\widehat{S}^3
$ 
depending on two dimensionless couplings
$\lambda$, $\kappa$ beyond the MSSM.  [Hatted (unhatted) capital letters denote
superfields (scalar superfield components).]  The associated trilinear soft terms are 
%
$
\lambda A_{\lambda} S H_u H_d + \frac{\kappa} {3} A_\kappa S^3 \,. 
$
%
The final two input parameters are 
%
$
\tan \beta = h_u/h_d$ and $\mu_\mathrm{eff} = \lambda
s \,, 
$
%
where $h_u\equiv
\vev {H_u}$, $h_d\equiv \vev{H_d}$ and $s\equiv \vev S$.
The Higgs sector of the NMSSM is thus described by the six parameters
$\lambda\ , \ \kappa\ , \ A_{\lambda} \ , \ A_{\kappa}, \ \tan \beta\ ,
\ \mu_\mathrm{eff}\ .$
In addition, values must be input for the gaugino masses 
and for the soft terms related to the (third generation)
squarks and sleptons that contribute to the
radiative corrections in the Higgs sector and to the Higgs decay
widths. 

The particle content of the NMSSM differs from
the MSSM by the addition of one CP-even and one CP-odd state in the
neutral Higgs sector (assuming CP conservation), and one additional
neutralino.  The result is three CP-even Higgs bosons ($h_{1,2,3}$)
two CP-odd Higgs bosons ($a_{1,2}$) and a total of
five neutralinos $\wtil\chi^0_{1,2,3,4,5}$.
 The NMHDECAY program \cite{Ellwanger:2004xm}, which
includes most LEP constraints, allows easy exploration
of Higgs phenomenology in the NMSSM.

In \cite{Dermisek:2005ar},
we found that the Next to Minimal Supersymmetric Model (NMSSM)
can avoid the fine-tuning  and hierarchy problems of the MSSM for
parameter choices that were consistent with all LEP
constraints available at the time. 
Defining the fine-tuning measure to be
\beq F={\rm Max}_p F_p \equiv {\rm
  Max}_p\left|{d\log \mz\over d\log p}\right|\,, 
\eeq 
where the parameters $p$ comprise all GUT-scale soft-SUSY-breaking
parameters, we found that $F<10$ could be achieved. Further, at
moderate $\tanb$, the parameter choices with such $F$ are always such
that the lightest CP-even Higgs boson, $\hi$, is SM-like as regards
its gauge and fermionic couplings, but decays primarily into a pair of
the lightest CP-odd Higgs bosons of the model, $\hi\to\ai\ai$.  The
importance of such decays was first emphasized in
\cite{Gunion:1996fb}, and later in \cite{Dobrescu:2000jt}, followed by
extensive work in
\cite{Ellwanger:2001iw,higsec3,Ellwanger:2005uu}.

We note that a light $a_1$ is natural in the NMSSM in the
$\kap\akap,\lam\alam\to 0$ limit. This can be understood as a
consequence of a global $U(1)_R$ symmetry of the scalar potential (in
the limit $\kap\akap, \lam\alam \to 0$) which is spontaneously broken
by the vevs, resulting in a Nambu-Goldstone boson in the
spectrum~\cite{Dobrescu:2000jt}.  This symmetry is explicitly broken
by the trilinear soft terms so that for small $\kap\akap, \lam\alam$
the lightest CP odd Higgs boson is naturally much lighter than other
Higgs bosons. For the $F<10$ scenarios, $\lam(\mz)\sim 0.15\div0.25$,
$\kap(\mz)\sim 0.15\div 0.3$, $|\akap(\mz)|<4\gev$ and
$|\alam(\mz)|<200\gev$, implying small $\kap\akap$
and moderate $\lam\alam$. The effect of $\lam\alam$ on $\mai$ is further
suppressed when the $\ai$ is largely singlet in nature, as is the case
for low-$F$ scenarios. Therefore, we always obtain small $\mai$.  We
note that small soft SUSY-breaking trilinear couplings at the
unification scale are generic in
SUSY breaking scenarios where SUSY breaking is mediated by the gauge
sector, as, for instance, in gauge or gaugino mediation.  Although the
value $\alam (m_Z)$ might be sizable due to contributions from gaugino
masses after renormalization group running between the unification
scale and the weak scale, $\akap$ receives only a small correction
from the running (such corrections being one loop suppressed compared
to those for $\alam$).  Altogether, a light, singlet $a_1$ is very
natural in models with small soft SUSY-breaking trilinear couplings at
the unification scale.  Finally, we note that the above $\lam(\mz)$
values are such that $\lam$ will remain perturbative when evolved up
to the unification scale, implying that the resulting
unification-scale $\lam$ values are natural in the context of model
structures that might yield the NMSSM as an effective theory below the
unification scale.

We will now discuss in more detail results for the NMSSM using the
representative fixed values of $\tanb=10$ and
$M_{1,2,3}(\mz)=100,200,300\gev$ while varying all other model
parameters. Similar results are obtained for other choices
of $\tanb$ and $M_{1,2,3}(\mz)$.  The points plotted for the NMSSM in
Fig.~\ref{zbblimits} show the $\cbb$ predictions for all parameter
choices in our scan that had $F<25$ and $\mai<2\mb$ and that
are consistent with the experimental and theoretical constraints
built into NMHDECAY as well as with limits from the preliminary LHWG
full analysis code.  The eight $F<10$
points are singled out. From Fig.~\ref{zbblimits} we see that these
latter points cluster near $\mhi\sim 98\div105\gev$ (see also Fig.~3
of \cite{Dermisek:2005ar}). We will see that
most are such that $\mhi$ and $\br(\hi\to b\anti b)$ are appropriate
for explaining the $\cbb$ excess. The other primary $\hi$ decay mode
for all the plotted points is $\hi\to \ai\ai$ with $\ai\to \tauptaum$
or light quarks and gluons (when $\mai<2\mtau$).  In
Table~\ref{eightpoints}, we give the precise masses and branching
ratios of the $\hi$ and $\ai$ for all the $F<10$ points.
We also give the number of standard deviations,
$\nobs$ ($\nexp$), by which the observed rate (expected rate obtained
for the predicted signal+background) exceeds the predicted background.
These are derived from $(1-CL_b)_{\rm observed}$ and $(1-CL_b)_{\rm
  expected}$ using the usual tables: e.g. $(1-CL_b)=0.32$, $0.045$,
$0.0027$ correspond to $1\sig$, $2\sig$, $3\sig$ excesses,
respectively. The quantity $s95$ is the factor by which the signal
predicted in a given case would have to be multiplied in order to exceed
the 95\% CL. All these quantities are obtained by processing each
scenario through the full preliminary LHWG confidence level/likelihood
analysis. 
If $\nexp$ is larger than $\nobs$ then the excess predicted by
the signal plus background Monte Carlo 
is larger than the excess actually observed and vice versa.
The points with $\mhi\lsim 100\gev$ have the largest $\nobs$.
Point 2 gives the best consistency between $\nobs$ and $\nexp$, 
with a predicted excess only
slightly smaller than that observed.  Points 1 and 3 also show
substantial consistency. For the 4th and
7th points, the predicted excess is only modestly larger
(roughly within $1\sigma$) compared to that observed. The 5th and 6th points are very
close to the 95\% CL borderline and have a predicted signal
that is significantly larger than the excess observed.   LEP is not
very sensitive to point 8. Thus, a significant
fraction of the $F<10$ points are very consistent with the observed
event excess.

\begin{table}
{\footnotesize
\hspace*{-.05in}
\begin{tabular}{|c|c|c|c|c|c|c|}
\hline
 $\mhi/\mai$ & \multicolumn{3}{c|} {Branching Ratios} &
$\nobs/\nexp$ & 
 $s95$  & $N_{SD}^{LHC}$ \cr
  (GeV) & $\hi\to b\anti b$  & $\hi\to \ai\ai$ &
$\ai\to\tau\anti\tau$ &  units of $1\sigma$ &  & \cr
\hline
\hline
\ 98.0/2.6 & 0.062 & 0.926 & 0.000 & 2.25/1.72  & 2.79 & 1.2 \cr   
100.0/9.3 & 0.075 & 0.910 & 0.852 & 1.98/1.88 & 2.40 & 1.5 \cr
100.2/3.1 & 0.141 & 0.832 & 0.000 & 2.26/2.78 & 1.31 & 2.5 \cr
102.0/7.3 & 0.095 & 0.887 & 0.923 & 1.44/2.08 & 1.58 & 1.6 \cr
102.2/3.6 & 0.177 & 0.789 & 0.814 & 1.80/3.12 & 1.03 & 3.3 \cr
102.4/9.0 & 0.173 & 0.793 & 0.875 & 1.79/3.03 & 1.07 & 3.6 \cr
102.5/5.4 & 0.128 & 0.848 & 0.938 & 1.64/2.46 & 1.24 & 2.4 \cr
105.0/5.3 & 0.062 & 0.926 & 0.938 & 1.11/1.52 & 2.74 & 1.2 \cr
\hline
\end{tabular}
}
\caption{Some properties of the $\hi$ and $\ai$ 
for the eight allowed points  with $F<10$ and $\mai<2\mb$ from our $\tanb=10$,
$M_{1,2,3}(\mz)=100,200,300\gev$ NMSSM scan.
The $\nobs$, $\nexp$ and $s95$  values are 
obtained after full processing of all $Zh$ final states
using the preliminary LHWG analysis code (thanks to
P. Bechtle). See text for details.
 $N_{SD}^{LHC}$ is the statistical
significance of the best 
``standard'' LHC Higgs detection channel for integrated luminosity
of $L=300\fbi$. }
\label{eightpoints}
\vspace*{-.24in}
\end{table}

We wish to emphasize that in our scan there are many, many points that
satisfy all constraints and have $\mai<2\mb$. The remarkable result is
that those with $F<10$ have a substantial probability that they
predict the Higgs boson properties that would imply a LEP $Zh\to
Z+b$'s excess of the sort seen. The smaller number of $F<10$ points
with $\mai$ substantially above $2\mb$ all predict a net $Z+b$'s
signal that is ruled out at better than $99\%$ CL by LEP data. Indeed,
all $F<25$ points have a net $h\to b$'s branching ratio, $\br(\hi\to
b\anti b)+\br(\hi\to \ai\ai\to b\anti b b\anti b)\gsim 0.85$, which is
too large for LEP consistency if $\mai$ is substantial.  Analysis of
points with $\mai$ very near $b\anti b$ decay threshold, but such that
$\ai\to b\anti b$ is dominant, is very subtle.  Such points arise for
$F<10$ and require further analysis in cooperation with the LHWG.

An important question is the extent to which the type of $h\to aa$
Higgs scenario (whether NMSSM or other) described here can be explored
at the Tevatron, the LHC and a future $\epem$ linear collider.  This
has been examined in the case of the NMSSM in
\cite{Gunion:1996fb,Ellwanger:2001iw,Ellwanger:2005uu}, with the
conclusion that observation of any of the NMSSM Higgs bosons may be
difficult at hadron colliders. At a naive level, the $\hi\to\ai\ai$
decay mode renders inadequate the usual Higgs search modes that might
allow $\hi$ discovery at the LHC.  Since the other NMSSM Higgs bosons
are rather heavy and have couplings to $b$ quarks that are not greatly
enhanced, they too cannot be detected at the LHC.  The last column of
Table~\ref{eightpoints} shows the statistical significance of the most
significant signal for {\it any} of the NMSSM Higgs bosons in the
``standard'' SM/MSSM search channels for the eight $F<10$ NMSSM
parameter choices. For the $\hi$ and $\ai$, the most important
detection channels are $\hi\to\gam\gam$, $W\hi+t\anti t\hi\to \gam\gam
\ell^{\pm}X$, $t\anti t \hi/\ai \to t\anti t b\anti b$, $b\anti b
\hi/\ai \to b\anti b \tauptaum$ and $WW\to \hi\to\tauptaum$ -- see
\cite{Ellwanger:2005uu}.  Even after $L=300\fbi$ of accumulated
luminosity, the typical maximal signal strength is at best $3.5\sigma$.
For the eight points of Table~\ref{eightpoints}, this largest signal
derives from the $W\hi+t\anti t\hi\to \gam\gam \ell^{\pm}X$ channel.
There is a clear need to develop detection modes sensitive to the
dominant $\hi\to \ai\ai\to \tauptaum\tauptaum$ decay channel.

Let us consider the possibilities.
One detection mode that can be considered is
$WW\to\hi\to\ai\ai\to 4\tau$.  
Second, recall that the  $\cntwo\to \hi \cnone$ channel provides a
signal in the MSSM when $\hi\to b\anti b$ decays are dominant. See,
for example, \cite{Marinelli:2004jp}.  It has not been studied for
$\hi\to\ai\ai\to 4\tau$ decays. If a light $\cnone$
provides the dark matter of the universe (as possible because of the
$\cnone\cnone\to \ai\to X$ annihilation channels for a light $\ai$,
see \cite{Gunion:2005rw,Belanger:2005kh} and references therein), the
$\mcntwo-\mcnone$ mass difference might be large enough to
allow such decays.  Diffractive production
\cite{Martin:2005rz}, $pp\to pp\hi\to pp X$, where the mass $\mx$ can
be reconstructed with roughly a $1-2\gev$ resolution, can
potentially 
reveal a Higgs peak, independent of the
decay of the Higgs.  A study \cite{hdiff} is underway to see if this
discovery mode works for the $\hi\to\ai\ai\to4\tau$ decay
mode as well as it appears to work for the simpler
SM $\hsm \to b\anti b$ case. The main issue may be whether events
can be triggered despite the soft nature of the decay products of the $\tau$'s 
present in $X$ when $\hi\to \ai\ai\to 4\tau$ 
as compared to $\hsm\to b\anti b$.

At the Tevatron it is possible that $Z\hi$ and $W\hi$ production, with
$\hi\to\ai\ai\to 4\tau$, will provide the most favorable
channels.  If backgrounds are small, one must simply accumulate enough
events. However, efficiencies for triggering on and isolating the
$4\tau$ final state will not be large. Perhaps one could also consider
$gg\to\hi\to\ai\ai\to 4\tau$ which would have
substantially larger rate. Studies are needed. If supersymmetry is
detected at the Tevatron, but no Higgs is seen, and if LHC discovery
of the $\hi$ remains uncertain, Tevatron studies of the $4\tau$ final
state might be essential. However, rates imply that the $\hi$ signal
could only be seen if Tevatron running is extended until $L>10\fbi$
has been accumulated.  Of course, the LHC {\it would} observe numerous
supersymmetry signals and {\it would} confirm that $WW\to WW$
scattering is perturbative, implying that something like a light Higgs
boson must be present, but direct detection of the $\hi$ might have to
rely on an extended Tevatron run.

Of course, discovery of the $\hi$
will be straightforward at an $\epem$ linear collider
via the inclusive $Z\h\to \ell^+\ell^- X$ reconstructed $M_X$
approach (which allows Higgs discovery independent
of the Higgs decay mode). Direct detection in the $Zh\to Z4\tau$
mode will also be possible. At a $\gam\gam$ collider, the
$\gam\gam\to \h\to 4\tau$ signal will
be easily seen \cite{Gunion:2004si}. 

In contrast, since (as already noted) the $\ai$ in these low-$F$ NMSSM scenarios is
fairly singlet in nature, its {\it direct} (i.e. not in $\hi$ decays) detection 
will be very challenging even at the ILC.  
Further, the low-$F$ points are all such that 
the other Higgs bosons are fairly heavy, typically above
$400\gev$ in mass, and essentially inaccessible at both the LHC and all
but a $\gsim 1\tev$ ILC.

We should note that much of the discussion above regarding Higgs
discovery is quite generic.  Whether the $a$ is truly the NMSSM
CP-odd $\ai$ or just a lighter Higgs boson into which
the SM-like $h$ pair-decays, hadron collider detection of the $h$
in its $h\to aa$ decay mode will be very challenging --- only
an $\epem$ linear collider can currently guarantee its discovery.

In conclusion, we reemphasize that the  prominent LEP event excess
in the $Z+b$'s channel for reconstructed Higgs mass of $\mh\sim 100\gev$
is consistent with a scenario in which 
the $ZZh$ coupling is SM-like but the $h$
decays mainly via $h\to aa\to \tauptaum\tauptaum$ (requiring $\ma<2\mb$)
leaving an appropriately reduced rate for $h\to b\anti b$. 
We strongly encourage the LEP groups to
push the analysis of the $Z4\tau$ channel in the hope of either ruling
out the $h\to aa\to 4\tau$ scenario, or finding a small
excess consistent with it. Either a positive or negative result would
have very important implications for Higgs searches at the Tevatron
and LHC.  Further, we have emphasized that the NMSSM models
with the smallest fine-tuning typically predict precisely the above
scenario with $h=\hi$ and $a=\ai$. We speculate that similar
results will emerge in other supersymmetric models with a Higgs sector
that is more complicated than that of the CP-conserving MSSM.

\begin{acknowledgments}
This work was supported by the U.S. Department of Energy
under grants DE-FG02-90ER40542 and DE-FG03-91ER40674.
JFG thanks the Aspen Center for Physics where part
of this work was performed. We are very grateful to 
P. Bechtle for processing our low-$F$ points through the full
preliminary LHWG
machinery and for a careful reading of and useful comments
on the paper. We thank A. Sopczak for
help in obtaining the $Z4b$ data files.  
R.D. thanks K. Agashe and N. Weiner for discussions.
\end{acknowledgments}

\end{document}